\documentclass[11pt]{article}

\usepackage[preprint]{acl}

\usepackage{times}
\usepackage{latexsym}

\usepackage[T1]{fontenc}
\usepackage[utf8]{inputenc}

\usepackage{microtype}

\usepackage{inconsolata}

\usepackage{graphicx}
\usepackage{amsmath}
\usepackage{booktabs}
\usepackage{multirow}
\usepackage{adjustbox}
\usepackage{array}
\usepackage{algorithm}
\usepackage{algorithmic}
\usepackage{fancyhdr}
\title{AI Chatbots as Professional Service Agents: Developing a Professional Identity}

\author{
  Wenwen Li$^1$, Kangwei Shi$^2$\thanks{Corresponding author}, Yidong Chai$^2$
\\
\\
 \textsuperscript{1}School of Management, Fudan University\\
 \textsuperscript{2}School of Management, Hefei University of Technology\\
 \href{mailto:liwwen@fudan.edu.cn}{liwwen@fudan.edu.cn,}
 \href{mailto:shikw@mail.hfut.edu.cn}{shikw@mail.hfut.edu.cn,}
 \href{mailto:chaiyd@hfut.edu.cn}{chaiyd@hfut.edu.cn}
}

\begin{document}
\maketitle

\begin{abstract}
    With the rapid expansion of large language model (LLM) applications, there is an emerging shift in the role of LLM-based AI chatbots from serving as general inquiry tools to acting as professional service agents. However, current studies often overlook a critical aspect of professional service agents: the act of communicating in a manner consistent with their professional identities. This is of particular importance in the healthcare sector, where effective communication with patients is essential for achieving professional goals, such as promoting patient well-being by encouraging healthy behaviors. To bridge this gap, we propose LAPI (LLM-based Agent with a Professional Identity), a novel framework for designing professional service agent tailored for medical question-and-answer (Q\&A) services, ensuring alignment with a specific professional identity. Our method includes a theory-guided task planning process that decomposes complex professional tasks into manageable subtasks aligned with professional objectives and a pragmatic entropy method designed to generate professional and ethical responses with low uncertainty. Experiments on various LLMs show that the proposed approach outperforms baseline methods, including few-shot prompting, chain-of-thought prompting, across key metrics such as fluency, naturalness, empathy, patient-centricity, and ROUGE-L scores. Additionally, the ablation study underscores the contribution of each component to the overall effectiveness of the approach.
\end{abstract}

\section{Introduction}
% \begin{figure}[ht]  % 'h' places the figure approximately here
%   \centering
%   \includegraphics[width=0.5\textwidth]{latex/figure/profff.pdf}  % 图片文件名是 fig1.png
%   \caption{Prompt used to assign weights to a question based on its relevance to the HBM categories.}  % 图片标题
%   \label{fig:prof}  % 图片标签
% \end{figure}

AI chatbots have gained great popularity worldwide to serve as professional assistants and provide services across various sectors\cite{dam-etal:llm-chatbots,xi-etal:llm-agents}. Integrating AI chatbots is a cost-effective strategy for organizations, delivering value to customers while reducing operational expenses. For instance, automation in customer interactions is projected to save retailers approximately \$439 billion annually by 2023, as they replace many human-operated customer service roles with chatbots \cite{williams:chatbots-retail-sales}. The emergent of generative AI, particularly large language models (LLMs), is revolutionizing the functionality of chatbots by significantly enhancing their interactive communication capabilities. LLMs, such as OpenAI's GPT-series and Llama, enabling chatbots to understand and generate natural language with remarkable fluency and context-awareness \cite{achiam-etal:gpt-4-technical-report,touvron-etal:llama-foundation-models}.

Recently, there is a trend of a significant transition in chatbot applications, shifting from general inquiry tools to professional service agents capable of delivering specialized support. The increasingly prevalent human-AI interactions in high-stake industries (e.g., healthcare and finance) require professional service agents to offer a more professional and functional experience than simple knowledge-enhanced inquiry tools\cite{safi-etal:chatbots-medical-applications}. This leads to a critical challenge in human-AI communication that has not been well addressed in existing studies, namely \textit{how AI chatbots can communicate with humans and answer questions in a manner consistent with the chatbots' professional identity}. Beyond simply responding to user queries, professional service agents always have a defined professional identity, such as that of a virtual doctor, which requires their responses to be aligned with professional goals, such as supporting patients' health behavior change (e.g., treatment or medication adherence and reducing substance abuse)\cite{clark-bailey:chatbots-healthcare}. However, existing AI chatbots primarily emphasize information retrieval and question answering, overlooking the professional intention and goals that are integral to professional fields \cite{lee-etal:ai-generated-medical-responses,singhal-etal:expert-level-medical-question-answering}. 

To develop a professional identity, an AI chatbot should have the ability to think and act like a professional. Firstly, the ability to analyze and break down complex professional tasks into more manageable components is essential. In contrast to mathematical and logical problems, which typically have clear chains of thought and definitive answers\cite{wei-etal:chain-of-thought}, many question-and-answer (Q\&A) tasks in a service scenario are open-ended and require free-form responses\cite{megahed-etal:misuse-of-ai-in-spc,singhal-etal:expert-level-medical-question-answering}. The challenge lies in effectively guiding these chatbots to adopt a professional thought process. Secondly, given their role as frontline professionals, AI chatbots must have the competence to effectively address user inquiries and deliver reliable services. However, it should be noted that LLMs are non-deterministic, meaning that the same input can generate different outputs, some of which may be incorrect or hallucinated\cite{farquhar-etal:detecting-hallucinations}. These drawbacks can result in occasional inaccurate answers or hallucination problems, which is fatal in the high-stakes field\cite{ji-etal:survey-hallucination-nlg}. Interactions with such chatbots have the potential to trigger undesirable behaviors in users, such as noncompliance among patients, which can negatively impact both service providers and users.

Motivated by to urgent need to provide professional service to users through LLM-based chatbots, our goal is to propose a novel framework, i.e., LAPI (LLM-based Agent with a Professional Identity), which aims to enhance the effectiveness of LLM-based chatbots in serving as professional service agents, particularly in the context of medical Q\&A services. To address the aforementioned challenges, we consider a subfield of linguistics, namely pragmatics, which how language is used for social interactions and the attainment of goals\cite{ciccia:cohesion-communication-burden} and the use of information from context to determine the effective means of achieving a given goal\cite{grice:logic-conversation}. From a pragmatic perspective, AI chatbots should provide useful responses and remove meaningless or incorrect information. 
The proposed approach involves two key steps: first, the comprehension and decomposition of complex professional tasks into manageable subtasks through a theory-guided task planning process; and second, the iterative updating and eventual generation of professional and ethical responses using a pragmatic entropy method. Specifically, we propose a theory-guided task planning process that employs a conceptual framework in the healthcare domain—namely, the Health Belief Model (HBM)—to guide AI chatbots in how they think and act to support interventions aimed at encouraging healthy behaviors. Furthermore, we propose a pragmatic entropy method that generates professional and ethical responses, with the objective of maximizing the clarity, empathy, and relevance of the chatbot’s outputs while ensuring alignment with ethical standards and professional goals.

In general, our contributions are:

\begin{itemize}
    \item We introduce a novel framework for developing a professional identity for an LLM-based chatbot, with the objective of employing the chatbot as a professional service agent for medical Q\&A services. Specifically, we propose a pragmatic entropy method for the generation of professional and ethical responses. Meanwhile, we provide an effective way to utilize a domain theory to guide task planning process.
    
    \iffalse
    \item We evaluate the components of our method, including the HBM-based task planning process, which breaks tasks into subtasks, and the pragmatic entropy method for generating professional responses. Effectiveness is assessed using both qualitative and quantitative metrics.
    \fi
    
    \item Experiments on various real medical Q\&A setting show that our method significantly improves response quality, providing more accurate, empathetic, and professional answers compared to baseline approaches.

\end{itemize}

\section{Research Background}

\subsection{Medical AI Chatbots}

The medical field represents a domain in which the implementation of AI chatbots is gaining traction as a means of facilitating access to information from the patient perspective and alleviating the workload on doctors\cite{safi-etal:chatbots-medical-applications}. 
A variety of methods and techniques have been employed in the development of medical chatbots for diverse objectives, including statistical methods, deep learning, and LLMs\cite{mihailidis-etal:coach-prompting,jothi-etal:humanoid-chatbot-medical,lee-etal:ai-generated-medical-responses}. These bots have been developed for diverse applications, including various aspects of health and medicine. These range from chatbots that support general health and activities of daily living to chatbots that assist specific disease management, such as heart disease and cancer\cite{wong-etal:health-conversational-system,lee-etal:ai-generated-medical-responses}. Prior to the advent of LLMs, conventional chatbots exhibited constrained contextual and language understanding, resulting in inaccurate responses and suboptimal interaction in a human-like manner. Recently, significant advancements have been made in the development of LLM-based medical chatbots, which are designed to facilitate human-like interactions with patients\cite{singhal-etal:expert-level-medical-question-answering}.

\subsection{Professional LLM-based Chatbots}

Chatbots have stepped into the age of LLM empowerment. The rapid development of LLMs enhances the capabilities of chatbots by improving their conversational proficiency and facilitating human-like interactions. ChatGPT exemplifies this category of chatbot, optimized to generate natural, human-like dialogue. Other widely used LLMs include GPT4, Llama, Mistral, ChatGLM2, and Gemma\cite{achiam-etal:gpt-4-technical-report,touvron-etal:llama-foundation-models,jiang-etal:mistral-7b,glm-etal:chatglm,gemma-etal:gemma-open-models}. Chatbots have been employed in customer service to address customer inquiries regarding products or services, either as an initial point of contact or as an alternative to speaking with a human representative\cite{hong-etal:expanding-chatbot-knowledge}. In addition to healthcare, LLM-based chatbots have gradually become powerful tools in areas such as retail, education, research, and many others\cite{megahed-etal:misuse-of-ai-in-spc,benzinho-etal:llm-chatbot-farm-to-fork,giudici-etal:home-automation-llm-chatbot}. To improve chatbot’s ability on professional tasks, such as programming education and home designing, various knowledge-enhanced strategies, such as fine-tuning, prompt engineering, and Retrieval-Augmented Generation (RAG), have been employed\cite{lewis-etal:retrieval-augmented-generation}.

\section{Theoretical Background of Professional Service Agents}

\subsection{Professional Service Agents}

From a philosophical standpoint, the term "agent" traditionally describes entities possessing desires, beliefs, intentions, and the capacity to action\cite{xi-etal:llm-agents,donald:actions-reasons-causes}. Human agents, for instance, are frequently employed by companies to handle customer service operations, focusing on addressing customer queries and ensuring satisfaction through effective communication and problem-solving skills. With the progression of AI, the concept of an "agent" has expanded to describe AI entities that are characterized by intelligent behavior, such as autonomy, proactiveness, and social capabilities. AI agents, particularly chatbots powered by large language models, have demonstrated remarkable versatility across various scenarios, attributable to their extensive inherent knowledge and human-like communication abilities\cite{huang-etal:agent-ai-holistic-intelligence,chong-etal:ai-chatbots-services-frontline}. 

LLM-based agents are typically structured around three key components. Agent Core serves as the main brain of the system, typically utilizing an LLM with general-purpose capabilities. Task Planning is responsible for decomposing complex tasks into manageable subtasks, thereby enhancing the agent's reasoning and reliability of responses. Actions of a chatbot are communication with humans. A professional AI agent's primary actions are aimed at high-level information transmission to fulfill a robot’s intent, a process referred to as “intention action.” A typical example of such agents is applications in the healthcare domain, where empathy-aware agents assist in diagnostics and knowledge retrieval to provide a user-friendly service\cite{mao-etal:biases-of-pretrained-models}.

\begin{figure}[h]  % 'h' places the figure approximately here
  \includegraphics[width=\columnwidth]{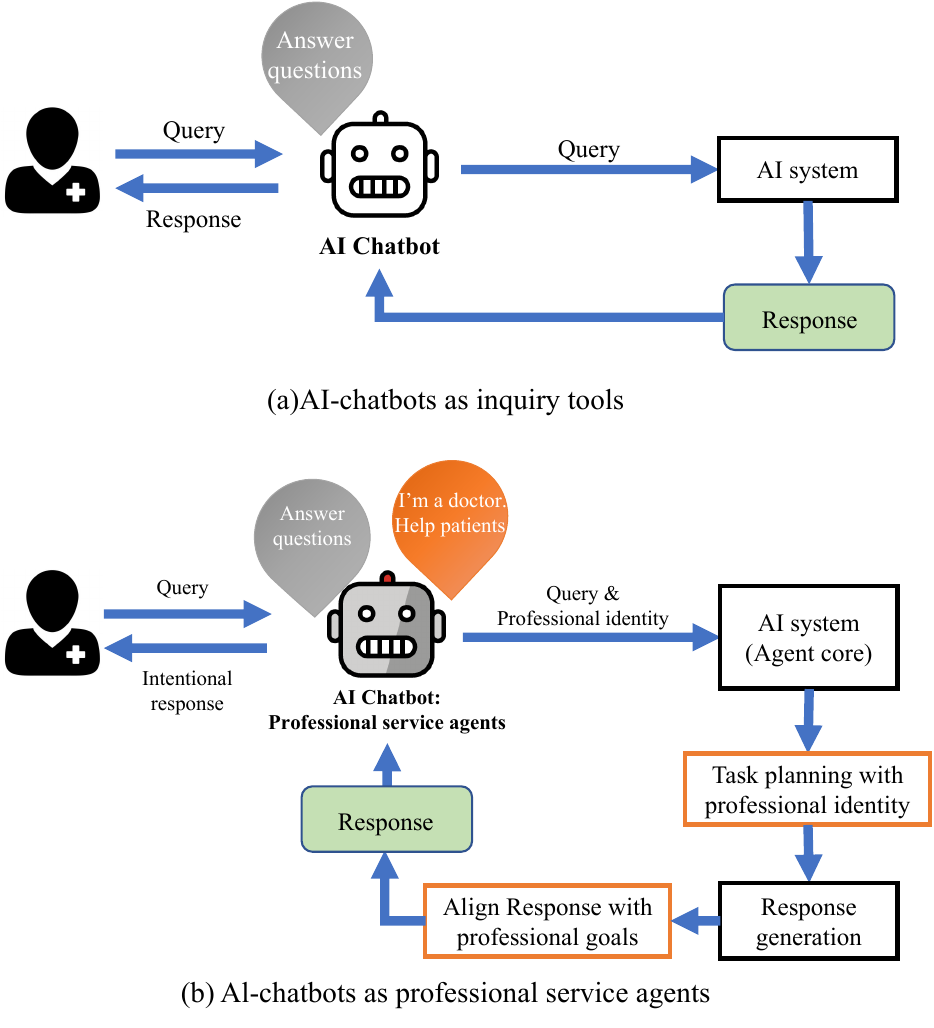}  % 图片文件名是 fig1.png
  \caption{Distinction Between AI Chatbots as Inquiry Tools and Professional Service Agents}  % 图片标题
  \label{fig:difference}  % 图片标签
\end{figure}

\begin{figure*}[t]  % 't' places the figure at the top of the page
  \centering
  \includegraphics[width=1\textwidth]{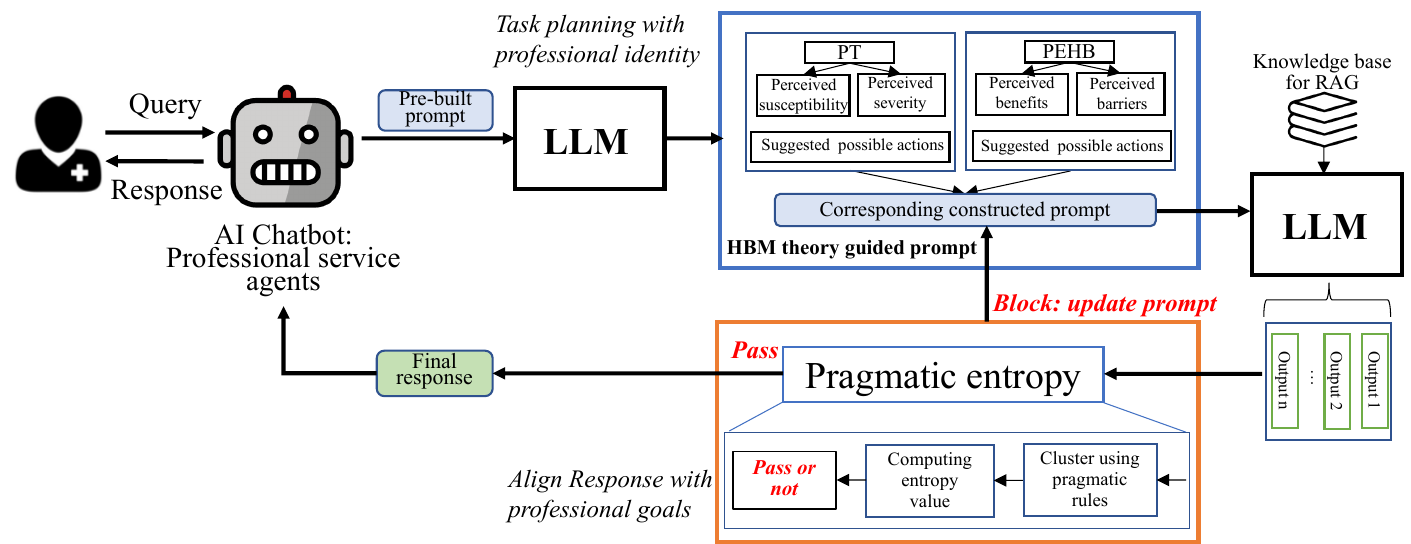}  % 图片文件名是 fig1.pdf
  \caption{Overview of LLM-based Agent with a Professional Identity (LAPI). The patient’s query is first assigned weights for the two HBM domains—Perceived Threat and Perceived Effectiveness of Health Behavior. An initial prompt is then generated based on these weights, followed by the addition of pragmatic rules. Multiple outputs are generated by the LLM, and pragmatic entropy optimization is applied to obtain the most appropriate response.}  % 图片标题
  \label{fig:framework}  % 图片标签
\end{figure*}

Responding to the trend of chatbots shifting from general inquire tools to professional service agents, we identify the significant distinction between AI-chatbots as inquiry tools and as professional service agents, as shown in Figure \ref{fig:difference}. Professional identity is a crucial attribute of a professional service agent, significantly shaping the agent’s goals, task planning, and action execution. Chatbots with professional identity have distinct professional intentions and goals, which in turn influence their task planning and action execution. Using medical Q\&A as an example, inquiry tools primarily focus on answering medical questions\cite{safi-etal:chatbots-medical-applications}. In contrast, professional service agents serve two purposes: responding to medical inquiries and promoting patient well-being by encouraging healthy behaviors, aligning with their professional identity (e.g., doctor). For task planning, the decomposition of complex professional tasks should adhere to established professional standards, informed by either theories or expert guidelines. Regarding actions, the chatbot's responses should align with professional goals. This alignment ensures that the chatbot not only addresses patient questions but also offers suggestions and persuasive guidance to foster healthy behaviors.

\subsection{Health Belief Model}

The Health Belief Model (HBM) is a widely used conceptual framework in health behavior research, which supports interventions aimed at encouraging healthy behaviors\cite{champion-skinner:health-belief-model}. Initially proposed in the 1950s, the model has evolved in response to practical public health concern and healthcare management. Rooted in a well-established body of psychological and behavioral theories, the HBM posits that health-related behavior depends primarily influenced by two factors: (1) the individual's desire to avoid illness or recover from it if already affected; and (2) the belief that a particular health action care prevent or mitigate the illness\cite{janz-becker:health-belief-model}.

The HBM comprises four key constructions: Perceived Susceptibility, Perceived Severity, Perceived Benefit, and Perceived Barrier. These constructs can be broadly categorized into two main aspects: Perceived Threat (including Perceived Susceptibility and Perceived Severity) and Perceived Effectiveness of Health Behavior (including Perceived Benefit and Perceived Barrier). Perceived Susceptibility refers to an individual’s belief about the likelihood of contracting a particular condition. Perceived Severity refers to the belief regarding the seriousness of contracting an illness or the consequences of leaving it untreated.  Perceived Benefit relates to the belief in the effectiveness of the recommended action to mitigate the disease threat. Perceived Barrier refers to the belief regarding the potential negative aspects or costs associated with the recommended health action. Individuals tend to weigh the anticipated benefits of an action against the perceived barriers, resulting in a belief in the effectiveness of the health behavior. These four constructions interact to guide individuals toward a preferred course of action.

The HBM has been employed to inform health intervention designs aimed at promoting healthy behaviors\cite{orji-etal:health-interventions-health-belief-model}, particularly as patient-centric (proactive) healthcare gains popularity. Previous research has shown that interventions grounded in established theories and models tend to be more effective than those developed based on intuition\cite{glanz:theory-at-a-glance}.

\section{Methods}

\subsection{Problem Formulation}

Given a patient's question \( q \) and a set of predefined pragmatic rules aligning with a professional identity \( R = \{r_1, r_2, r_3, \dots, r_m\} \), the objective is to generate a response \( s \) that not only provides an professional answer to the query but also adheres to these established rules, thereby reflecting the professional identity (i.e., the doctor in our experiment setting). The desired response \( s \) should satisfy the following condition: it adheres to the pragmatic rules with a minimal level of uncertainty in the response distribution. The problem can be formally expressed as:

\begin{equation}
    s = \text{LLM}(q, R, \tau, \epsilon)
\end{equation}

where \( \tau \) is the satisfaction rate threshold, and \( \epsilon \) is the pragmatic entropy threshold. 

Figure~\ref{fig:framework} illustrates the proposed framework, i.e., LLM-based Agent with a Professional Identity (LAPI), which includes a theory-guided task planning process and a pragmatic entropy method. In the following sections, we introduce details on our specific implementation of these two components. 

\subsection{Theory-guided Task Planning}

We employ the Health Belief Model (HBM) as a guiding framework to structure our design, decomposing the professional goal of promoting healthy behavior into manageable sub-tasks. These sub-tasks include: fostering individuals’ accurate perception of disease risk, aiding individuals in developing a realistic understanding of the consequences associated with a condition or recommended action, enhancing individuals' awareness of the potential benefits of adopting healthy behaviors, and offering strategies to overcome perceived barriers to implementing these behaviors.

The task planning process begins by assessing the relevance of the user’s query to the two core categories of HBM: \textit{Perceived Threat} and \textit{Perceived Effectiveness of Health Behavior}. To achieve this, we first prompt the LLM to estimate the proportions of the query related to these two categories. 

% \begin{figure}[h]  % 'h' places the figure approximately here
%   \centering
%   \includegraphics[width=0.5\textwidth]{latex/figure/prompt.pdf}  % 图片文件名是 fig1.png
%   \caption{Prompt used to assign weights to a question based on its relevance to the HBM categories.}  % 图片标题
%   \label{fig:prompt}  % 图片标签
% \end{figure}

Based on the prompt shown in Figure \ref{fig:prompta1} of Appendix \ref{sec:appendixa}, the LLM assesses the query's relevance to each category and assigns a corresponding weight to each. Next, we generate the initial prompt \( p_{\text{initial}} \) for the LLM by constructing two specific prompts: one fully focused on the \textit{Perceived Threat} category, and another fully focused on \textit{Perceived Effectiveness of Health Behavior}. These prompts are pre-configured to ensure that the LLM generates an initial prompt that accurately reflects the correct placement of the question within the HBM theory framework. 
The content of the prompt is detailed in Figure \ref{fig:prompta2} of Appendix \ref{sec:appendixa}.
%The detailed content of these prompts is provided in Appendix A.

% Based on the prompt in the left section of Figure~\ref{fig:prompta1} of Appendix~\ref{sec:appendixa}, the LLM evaluates the query's relevance to each Health Belief Model (HBM) category and assigns corresponding weights. Subsequently, an initial prompt, \( p_{\text{initial}} \), is generated using two tailored prompts: one focused exclusively on the \textit{Perceived Threat} category and another on the \textit{Perceived Effectiveness of Health Behavior} category. These pre-configured prompts ensure accurate alignment with the HBM framework.
% The prompt content is detailed in the right section of Figure~\ref{fig:prompta1} in Appendix~\ref{sec:appendixa}.

\subsection{Pragmatic Entropy}
Pragmatic entropy aims to assist the AI chatbot in generating professional and ethical responses by reducing the entropy of model outputs across different pragmatic rule categories and maximizing the proportion of responses that satisfy all predefined pragmatic rules \( R = \{r_1, r_2, r_3, ..., r_m\} \).

Given a prompt \( p \), the LLM generates a set of \( n \) responses by sampling from its output distribution with temperature 1, which enables the model's outputs to more authentically reflect its internal representations and reasoning processes:

\begin{equation}
    S = \text{LLM}(p, \text{temperature}=1)
\end{equation}

where \( S = \{s_1, s_2, \dots, s_n\} \) represents the generated responses.

Each response \( s \in S \) is classified into categories \( C_j \) based on its satisfaction of specific subsets of the pragmatic rules \( R  \). This classification process is judged by the LLM itself. The following metrics are computed to evaluate the quality of the prompt.

The proportion of responses satisfying all rules is given by:
    \begin{equation}
    P(C_\text{all}|S) = \frac{\left| \{s_i \in S \mid s_i \in C_\text{all} \} \right|}{n}
    \end{equation}
    where \( \left| \{s_i \in S \mid s_i \in C_\text{all} \} \right| \) denotes the number of responses in \( S \) that fall into category \( C_\text{all} \), and \( n \) is the total number of generated responses.

The entropy of the output distribution across all categories \( C_j \):
    \begin{equation}
        PE(S) = - \sum_{C_j} P(C_j|S) \log P(C_j|S)
    \end{equation}
    where \( P(C_j|S) = \frac{\left| \{s_i \in S \mid s_i \in C_j \} \right|}{n} \) represents the proportion of responses in \( S \) that fall into \( C_j \). 

To refine the prompt iteratively, the process consists of generating an improved prompt, evaluating it quantitatively, and iterating until either the desired criteria are met or the maximum number of iterations is reached.

To improve the prompt, the current prompt \( p \) along with information on unmet conditions (\textit{Ucs}) is fed into the large language model to generate a refined version \( p_{\text{new}} \):
\begin{equation}
    p_{\text{new}} = \text{LLM}(p, Ucs, \text{temperature}=0.3)
\end{equation}
where \( p_{\text{new}} \) is the updated prompt. \textit{Ucs} refer to the missing pragmatic rules identified within the category that has the largest proportion in categories that have not satisfied all the rules. The parameter \( \text{temperature}=0.3 \)\cite{moslem2023finetuninglargelanguagemodels,wassie2024machinetranslationgeezlanguage} is chosen to balance stability and flexibility in the prompt generation process.

% The refined prompt \( p_{\text{new}} \) is evaluated quantitatively by generating a new set of \( n \) responses:
% \begin{equation}
%     S_{\text{new}} = \text{LLM}(p_{\text{new}}, \text{temperature}=1)
% \end{equation}
% Each response \( s \in S_{\text{new}} \) is classified into categories \( C_j \), and the satisfaction rate \( P(C_\text{all}|S_{\text{new}}) \) and pragmatic entropy \( PE(S_{\text{new}}) \) are recomputed.

If the satisfaction rate \( P(C_\text{all}|S) \) meets or exceeds a predefined threshold \( \tau \), and the pragmatic entropy \( PE(S) \) is below a threshold \( \epsilon \), the prompt \( p \) is considered optimal, and the process terminates with \( p^* = p \). Otherwise, the model proceeds to refine the prompt. The overall process can be seen in Algorithm \ref{alg:pragmatic_entropy}.

\begin{algorithm}[H]
\caption{Pragmatic entropy}
\label{alg:pragmatic_entropy}
\begin{algorithmic}[0]
\STATE \textbf{Input:} Current prompt \( p \), pragmatic rules \( R \), satisfaction threshold \( \tau \), entropy threshold \( \epsilon \), maximum iterations \( T \).
\STATE \textbf{Output:} Final response \( s^* \).
\end{algorithmic}
\begin{algorithmic}[1]
\STATE Set iteration counter \( t \gets 0 \);
\WHILE{\( t < T \)}
    \STATE Generate response set \(S\) using Eq.(2);
    \STATE Compute \( P(C_\text{all}|S)\) using Eq.(3);
    \STATE Compute \( PE(S)\) based on Eq.(4);
    \STATE Select \( s^* \) randomly from \( C_\text{all}\);
    \IF{\( P(C_\text{all}|S) \geq \tau \) \textbf{and} \( PE(S) \leq \epsilon \)}
        \STATE \textbf{Return} \( s^* \);
    \ELSE
        \STATE Identify \( Ucs \) from categories that have not satisfied all the rules;
        \STATE Get prompt \(p_{\text{new}}\) using Eq.(5);
        \STATE Set \( p \gets p_{\text{new}} \);
    \ENDIF
    \STATE Increment iteration counter \( t \gets t + 1 \)
\ENDWHILE
\STATE \textbf{Return} \( s^* \)
\end{algorithmic}
\end{algorithm}

\section{Experiments}
\subsection{Experimental Settings}

\begin{table*}[t]
\small
\centering
\renewcommand{\arraystretch}{0.8} % Adjust row height for better readability
\begin{adjustbox}{center,width=\textwidth}
\begin{tabular}{cc|cccccccc}
\toprule
\textbf{Models} & \textbf{Approach} & \textbf{Flu.} & \textbf{Coh.} & \textbf{Nat.} & \textbf{Emp.} & \textbf{Pat-C.} & \textbf{Rec.} & \textbf{Pre.} \\ \midrule
\multirow{5}{*}{Llama-2-7B-Chat} 
& Zero-Shot                     & 3.90  & 3.60  & 2.95 & 2.10  & 2.45 & 0.39 & 0.07 \\ 
& Zero-Shot + Pragmatic rules    & 4.00  & 3.50  & \textbf{3.40}  & 2.70  & 3.00  & 0.33 & 0.10 \\ 
& RAG-based Few-Shot            & 3.60  & 2.50  & 2.65 & 2.20  & 2.45 & 0.38 & 0.07 \\ 
& Chain-of-Thought              & 4.00  & 3.60  & 2.80  & 2.20  & 2.40  & \textbf{0.40} & 0.07 \\ 
& LAPI       & \textbf{4.00}  & \textbf{3.60}  & 3.05 & \textbf{2.90}  & \textbf{3.10}  & 0.35 & \textbf{0.07} \\ \midrule
\multirow{5}{*}{Llama-2-13B-Chat} 
& Zero-Shot                     & 3.75 & 3.60 & 2.85 & 2.25 & 2.65 & 0.38 & 0.08 \\ 
& Zero-Shot + Pragmatic rules    & 4.00  & 3.75 & 3.10  & 3.05 & 3.25 & 0.33 & 0.12 \\ 
& RAG-Based Few-Shot            & 3.60  & 2.65 & 2.75 & 2.35 & 2.25 & 0.31 & 0.07 \\ 
& Chain-of-Thought              & 3.65 & 3.65 & 2.95 & 2.10  & 2.50  & \textbf{0.39} & 0.07 \\ 
& LAPI                         & \textbf{4.00}  & \textbf{3.75} & \textbf{3.15} & \textbf{3.40}  & \textbf{3.45} & 0.36 & \textbf{0.07} \\ \midrule

\multirow{5}{*}{Llama-2-70B-Chat} 
& Zero-Shot                     & 3.90  & 3.45 & 3.15 & 2.15 & 2.80  & 0.35 & 0.09 \\ 
& Zero-Shot + Pragmatic rules    & 4.00  & 3.75 & 3.40  & 2.80  & 3.35 & 0.35 & 0.08 \\ 
& RAG-Based Few-Shot            & 3.80  & 3.20  & 2.95 & 2.35 & 2.95 & 0.31 & 0.11 \\ 
& Chain-of-Thought              & 3.95 & 3.45 & 3.05 & 2.40  & 2.70  & \textbf{0.38} & 0.07 \\ 
& LAPI                         & \textbf{4.00}  & \textbf{3.75} & \textbf{3.50}  & \textbf{3.35} & \textbf{3.35} & 0.36 & \textbf{0.07} \\ \midrule

\multirow{6}{*}{GPT-3.5-Turbo} 
& Zero-Shot                     & 4.00  & 3.60  & 3.30  & 2.35 & 3.15 & 0.25 & 0.15 \\ 
& Zero-Shot + Pragmatic rules    & 4.00  & 3.60  & 3.45 & 3.15 & 3.55 & 0.26 & 0.16 \\ 
& RAG-Based Few-Shot            & 4.00  & 3.45 & 3.25 & 2.10  & 3.30  & 0.26 & 0.24 \\ 
& Chain-of-Thought              & 4.00  & 3.70  & 3.10  & 2.30  & 3.10  & 0.29 & 0.16 \\ 
& Semantic Entropy              & 4.00  & 3.40  & 3.10  & 2.30  & 3.15 & 0.21 & 0.21 \\ 
& LAPI                         & \textbf{4.00}  & \textbf{3.85} & \textbf{3.60} & \textbf{3.50}  & \textbf{3.70} & \textbf{0.29} & \textbf{0.13} \\ \midrule

\multirow{6}{*}{GPT-4} 
& Zero-Shot                     & 4.00  & 3.70  & 3.05 & 2.10  & 2.90  & 0.35 & 0.08 \\ 
& Zero-Shot + Pragmatic rules    & 4.00  & 3.90  & 3.50  & 3.55 & 3.80  & 0.35 & 0.08 \\ 
& RAG-Based Few-Shot            & 4.00  & 3.65 & 3.25 & 2.30  & 3.25 & 0.36 & 0.11 \\ 
& Chain-of-Thought              & 4.00  & 3.70  & 3.10  & 2.35 & 2.85 & 0.37 & 0.08 \\ 
& Semantic Entropy              & 4.00  & 3.65 & 3.05 & 2.40  & 2.80  & 0.35 & 0.07 \\ 
& LAPI                         & \textbf{4.00}  & \textbf{3.90}  & \textbf{3.95} & \textbf{3.85} & \textbf{3.80}  & \textbf{0.38} & \textbf{0.07} \\ 

\bottomrule
\end{tabular}
\end{adjustbox}
\caption{Evaluation results. The table presents the performance of different prompting approaches across various evaluation metrics. Flu., Coh., Nat., Emp., and Pat-C. represent Fluency, Coherence, Naturalness, Empathy, and Patient-Centricity, respectively. Rec. and Pre. refer to ROUGE-L Recall and Precision. Higher scores for the first six metrics indicate better performance, while a lower score for Pre. reflects a more desirable outcome. \textbf{Bold}: Best-performing paradigm per metric.}
\end{table*}

\begin{figure*}[t]  % 't' places the figure at the top of the page
  \centering
  \includegraphics[width=1\textwidth]{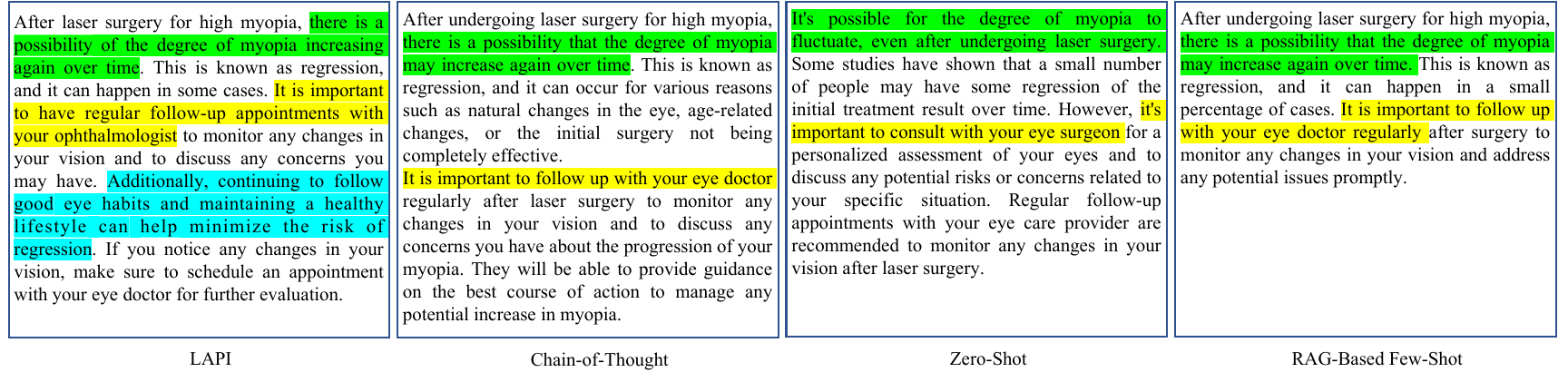}  % 图片文件名是 fig1.pdf
  \caption{Overview of GPT-3.5-Turbo's responses using different methods to answer the patient query: "After undergoing laser surgery for high myopia, will the degree of myopia increase again?"} % 图片标题
  \label{fig:casestudy}  % 图片标签
\end{figure*}

\paragraph{Dataset and Metrics} 
The dataset used in our experiments was collected from a reputable ophthalmology hospital and comprises 400 real-world patient inquiries paired with concise and professional answers provided by expert ophthalmologists. To support Retrieval-Augmented Generation (RAG), we utilized 300 patient-question-answer pairs as the knowledge base for retrieval. To assess the efficacy of the proposed method, a test set of 10 patient inquiries was randomly selected. This test set was used to evaluate the performance of the method in handling real-world medical question-and-answer scenarios. 

To comprehensively evaluate the performance of our proposed method, we utilized two categories of metrics: LLM-based and non-LLM-based metrics. For LLM-based evaluation, we employed G-EVAL\cite{liu-etal-2023-g}, a state-of-the-art metric designed to assess various qualitative aspects of large language model outputs, including Fluency, Coherence, Naturalness, Empathy, and Patient-Centricity, where higher scores indicate better performance. Detailed descriptions of these five evaluation met- rics can be found in Appendix B.  G-EVAL has been shown to exhibit strong consistency with human evaluations\cite{zhu-etal-2025-factual,gao2024largelanguagemodelswikipediastyle}, making it a reliable tool for qualitative assessment. For non-LLM-based evaluation, we adopted ROUGE-L recall and precision\cite{lin-2004-rouge} as objective n-gram matching-based metrics. ROUGE-L Recall measures the extent to which the generated responses cover the content of the doctor-provided answers, with higher values indicating better coverage, while ROUGE-L Precision evaluates the informativeness of the generated responses, with lower values indicating that the generated responses contain more information.

% ROUGE-L Recall measures the extent to which the generated responses cover the content of the doctor-provided answers, with higher values indicating better coverage, while ROUGE-L Precision evaluates the informativeness of the generated responses, with lower values indicating response contains more information. 
% Detailed descriptions of these five evaluation metrics can be found in Appendix \ref{sec:appendixeval}.

\paragraph{Baselines}
We categorize the baseline methods into two groups: prompt design-based methods and quality inspection-based methods. For the prompt design-based methods, we include the following: \textbf{Zero-Shot}, where the model generates responses directly without additional guidance; \textbf{Zero-Shot + Pragmatic Rules}, which enhances the zero-shot approach by incorporating explicit pragmatic rules tailored to professional contexts; \textbf{RAG-based Few-Shot}\cite{ram-etal-2023-context,yu-etal-2023-retrieval}, a retrieval-augmented generation strategy that retrieves relevant examples from a knowledge base to assist response generation; and \textbf{Chain-of-Thought (CoT)}\cite{kojima2023largelanguagemodelszeroshot}, which encourages the model to generate step-by-step reasoning paths to improve response accuracy and relevance. For the quality inspection-based methods, we employed \textbf{Semantic Entropy}\cite{farquhar-etal:detecting-hallucinations}, a technique designed to evaluate the confidence of the model regarding specific knowledge points in its generated responses. The confidence scores are then fed back into the model, prompting it to revise and improve its outputs.

\paragraph{Experiment Details}
We evaluate our proposed method on five LLMs: Llama-2-7B-Chat, Llama-2-13B-Chat, Llama-2-70B-Chat \cite{touvron2023llama2openfoundation}, GPT-3.5-Turbo (gpt-3.5-turbo-0125), and GPT-4 (gpt-4-0125-preview), selected for their diverse scales and capabilities, representing both open-source and closed-source models. To enhance the alignment of responses with professional and human-like communication, we incorporated pragmatic rules developed through a streamlined process: we reviewed medical literature, identified key communication principles (focusing on reply format, information content, and user emotional impact), and distilled these into actionable rules, which are detailed in Appendix \ref{sec:appendixrules}.

 For the RAG-based Few-Shot method, we employed a retrieval-augmented generation strategy using a BERT encoder to retrieve the three most relevant question-answer pairs based on cosine similarity, which were then used as prompts to guide response generation. In the LAPI method, we set \( n \) to 5, \( \tau \) to 0.5, and \( \epsilon \) to 0.8. For the Semantic Entropy method, due to the high text comprehension capability required for this method, which was only applied to GPT models.
 
\subsection{Experimental Results}

Our proposed approach LAPI demonstrates superior performance across various models, as evidenced by the results in Table 1. From the results perspective, LAPI consistently outperforms other methods in terms of Fluency, Coherence, Naturalness, Empathy, Patient-Centricity, ROUGE-L Recall, and Precision. For instance, in the same model family (e.g., Llama-2-7B-Chat), LAPI provides the highest scores across multiple metrics highlighting the effectiveness of LAPI in generating more precise, empathetic, and coherent responses, tailored to professional service agent tasks. From the model perspective, we observe that LAPI performs even better on larger models such as GPT-3.5-Turbo and GPT-4. The enhanced textual comprehension capabilities of these models enable them to generate more accurate prompts and subsequently produce better results. LAPI benefits from this ability, yielding more contextually rich and reliable outputs. 

The combination of pragmatic rules and the HBM framework allows LAPI to produce responses that are not only accurate and fluent but also deeply empathetic and patient-centric. By integrating step-by-step reasoning and continuous feedback mechanisms, our method ensures the delivery of high-quality, professional, and reliable responses. These results demonstrate the potential of LAPI to improve interactions in professional service applications, especially in the healthcare sector where empathy and clarity are paramount.

\subsection{Ablation Study}

We conducted an ablation study using GPT-3.5-Turbo as the base model to evaluate the effectiveness of each component in our proposed method. Table 2 presents the results of the ablation study, where we compare three different setups: the full LAPI method, the model without the HBM framework w/o HBM, and the model without the Pragmatic Entropy component w/o PE. As shown in the table, the full LAPI method outperforms both variants in all evaluation metrics, including Fluency, Coherence, Naturalness, Empathy, Patient-Centricity, ROUGE-L Recall, and Precision.

\begin{table}[h]
\small
\centering
\label{tab:ablation-study}
\resizebox{\columnwidth}{!}{
\renewcommand{\arraystretch}{1}
\begin{tabular}{l|ccccccc}
\toprule
\textbf{Approach} & \textbf{Flu.} & \textbf{Coh.} & \textbf{Nat.} & \textbf{Emp.} & \textbf{Pat-C.} & \textbf{Rec.} & \textbf{Pre.} \\ \midrule
LAPI   & \textbf{4.00} & \textbf{3.85} & 3.60 & \textbf{3.50}  & \textbf{3.70}  & \textbf{0.29}   & \textbf{0.13} \\ 
w/o HBM & 4.00 & 3.70  & \textbf{3.75} & 3.30 & 3.60 & 0.27   & 0.15 \\
w/o PE & 3.90  & 3.70  & 3.50 & 2.95 & 3.60 & 0.26   & 0.16 \\

\bottomrule
\end{tabular}}

\caption{Ablation study on GPT-3.5-turbo.}
\end{table}

The results indicate that each component contributes positively to the overall performance. Specifically, removing the HBM framework (w/o HBM) leads to a slight decrease in the Coherence and Empathy metrics, while the absence of the Pragmatic Entropy (w/o PE) reduces the overall Naturalness and Empathy. These findings suggest that combining both the HBM framework and Pragmatic Entropy is more effective than using them separately, as the combined approach leverages the strengths of both components, resulting in more fluent, coherent, and empathetic responses tailored to professional service applications.

% \begin{table}[h]
% \small
% \centering
% \label{tab:ablation-study}
% \renewcommand{\arraystretch}{1.2}
% \begin{tabular}{l|ccccccc}
% \toprule
% \textbf{Approach} & \textbf{Flu.} & \textbf{Coh.} & \textbf{Nat.} & \textbf{Emp.} & \textbf{Pat-C.} & \textbf{Rec.} & \textbf{Pre.} \\ \midrule
% LAPI             & \textbf{4.00} & \textbf{3.85} & 3.60          & \textbf{3.50}  & \textbf{3.70}  & \textbf{0.29} & \textbf{0.13} \\
% w/o HBM          & 4.00          & 3.70          & \textbf{3.75} & 3.30          & 3.60          & 0.27          & 0.15          \\
% w/o PE           & 3.90          & 3.70          & 3.50          & 2.95          & 3.60          & 0.26          & 0.16          \\
% \bottomrule
% \end{tabular}
% \caption{Ablation study on GPT-3.5-turbo.}
% \end{table}

\subsection{Case Study}

To further illustrate the capabilities of our proposed method, we present a case study using GPT-3.5-Turbo to answer the patient query.
Figure \ref{fig:casestudy}  provides a comparison of the responses generated by four different prompting strategies: LAPI, CoT, Zero-Shot, and RAG-Based Few-Shot. In the responses, green text directly answers the question, yellow text promotes regular doctor contact, and blue text provides actionable patient advice.
As observed in the case study, LAPI (left) outperforms others, providing a more comprehensive and detailed response that directly answers the question, stresses the importance of regular ophthalmologist follow-ups, and offers advice on healthy eye habits. This showcases the benefit of integrating the HBM framework and Pragmatic Entropy. In contrast, other methods provide direct answers and some patient advice but lack LAPI's level of detail and actionable guidance. For instance, CoT misses lifestyle advice, and RAG-Based Few-Shot doesn't emphasize continuous patient-doctor communication as strongly as LAPI. 

The LAPI framework also demonstrates superior performance compared to various CoT variants, with detailed results presented in Appendix \ref{sec:cotvar}.

% As observed in the case study, LAPI (on the left) provides a more comprehensive and detailed response compared to other methods. Not only does it provide a direct and clear answer to the question, but it also emphasizes the importance of maintaining regular follow-up appointments with an ophthalmologist and provides advice on healthy eye habits. This highlights the advantage of incorporating both the HBM framework and Pragmatic Entropy, as it allows the model to generate a richer set of responses that prioritize patient well-being.

% In contrast, while methods like Chain-of-Thought and Zero-Shot also offer a direct answer and some degree of patient advice, they lack the same level of detail and actionable guidance. For example, the response generated by Chain-of-Thought suggests regular follow-up appointments but doesn't include advice on lifestyle factors, which is included in the LAPI response. Additionally, RAG-Based Few-Shot also provides a helpful answer but does not emphasize as much on continuous patient-doctor communication as LAPI does.

% This case study demonstrates the effectiveness of our method in generating more comprehensive, empathetic, and actionable responses, which are essential for professional service agents, particularly in sensitive domains like healthcare.

\section{Conclusion}
% In this work, we introduce a novel approach for designing LLM-based professional service agents tailored for medical Q\&A services. 
% Our method ensures communication that aligns with professional identities by combining theory-guided task planning and a pragmatic entropy approach to generate ethical and patient-centric responses. 
% Experiments show that our method outperforms baseline techniques in fluency, empathy, patient-centricity, and ROUGE-L scores. This research highlights the potential of enhancing AI chatbot communication, particularly in healthcare, and encourages further development of more effective methods in future work.
In this work, we propose a novel approach for designing LLM-based professional service agents specifically tailored for medical Q\&A services. Our method promotes communication aligned with professional identities by integrating theory-guided task planning and a pragmatic entropy method to generate ethical, coherent, and patient-centric responses. 
Experimental results demonstrate that our approach outperforms baseline methods in fluency, empathy, patient-centricity, and ROUGE-L scores. This study underscores the potential of enhancing AI chatbot communication in healthcare and encourages the continued development of more effective methods in future research.

\section{Limitations}
While our LAPI framework shows significant promise for enhancing professional communication in medical Q\&A services, several aspects require further refinement. The current evaluation primarily relies on large language models to assess output quality, which aligns closely with human judgment. However, incorporating real-world testing, such as deploying the model in clinical settings, would further validate its practical effectiveness and applicability. While this didn’t affect the validity of the proposed method—since results across different LLM sizes showed notable improvements over baseline methods in attributes like Empathy—the smaller sample size might limit how broadly the findings can be applied. Furthermore, the experiments focused solely on medical Q\&A scenarios. While the framework is designed to apply to other domains like law, education, or finance, its effectiveness in these areas requires further validation.

\section{Ethical statements}
The dataset was provided by medical professionals from a reputable ophthalmology hospital and consists solely of anonymized question-answer pairs without any identifiable patient information. During the data collection process, patients provided verbal consent for their questions to be used for academic research purposes and were advised to avoid including personal or identifiable information in their queries. This process ensured that no personal health information was collected or used. Furthermore, the dataset was reviewed and approved by the Institutional Review Board (IRB) of the collaborating ophthalmology hospital prior to its use in our study, ensuring compliance with ethical standards.

\section{Acknowledgements}
We would like to thank all the reviewers for their helpful feedback, and EMNLP 2025 and ACL Rolling Review organizers for their efforts.
This work was supported by the grants from National Natural Science Foundation of China (No.72201068, 72131004, 72342011).

% Second, our evaluation primarily leverages LLM-based metrics, which align well with human judgment but may benefit from supplementary real-world validation; subsequent studies could deploy the model in clinical settings to further confirm its practical utility and robustness.


%\bibliography{acl_latex}

\clearpage
\appendix

\section{Prompts Used in Theory-guided Task Planning}
\label{sec:appendixa}

\begin{figure}[h]  % 'h' places the figure approximately here
  \centering
  \includegraphics[width=\columnwidth]{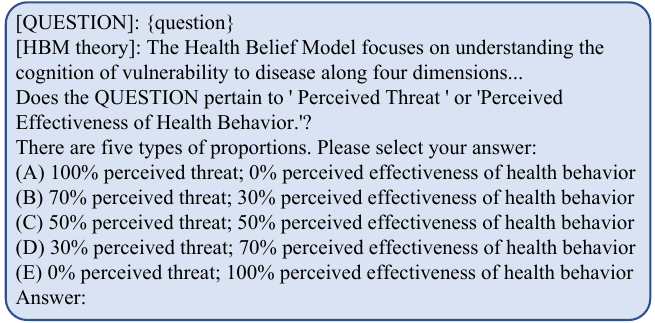}  % 图片文件名是 fig1.png
  \caption{Prompt used to assign weights to a question based on its relevance to the HBM categories.}  % 图片标题
  \label{fig:prompta1}  % 图片标签
\end{figure}

\begin{figure}[h]  % 'h' places the figure approximately here
  \centering
  \includegraphics[width=\columnwidth]{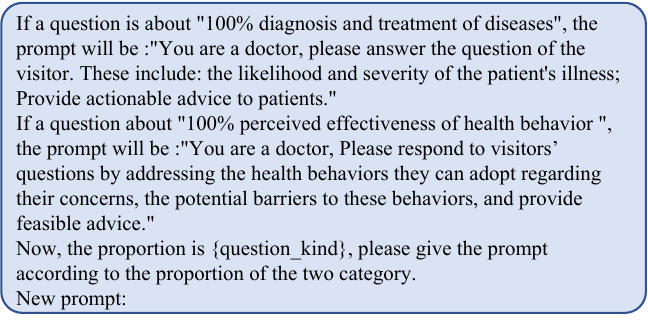}  % 图片文件名是 fig1.png
  \caption{Prompts used to generate an initial prompt that accurately reflects the correct placement of the question within the HBM theory framework.}  % 图片标题
  \label{fig:prompta2}  % 图片标签
\end{figure}

\section{Evaluation Metrics}
\label{sec:appendixeval}

In the following content, we provide a detailed description of the five key evaluation metrics used to assess the performance of the proposed method, all of which are evaluated using the G-EVAL\cite{liu-etal-2023-g} framework with GPT-4. This approach ensures a comprehensive and consistent evaluation of the generated responses.

\begin{itemize}
    \item \textbf{Fluency (1--4 points):} Quality of the response in terms of grammar, spelling, punctuation, word choice, and sentence structure.
    \begin{itemize}
        \item \emph{4 points:} No noticeable issues with grammar, spelling, punctuation, word choice, and sentence structure; the response is concise and clear.
        \item \emph{3 points:} The response has few or no grammatical or spelling errors; some word choices or structures may seem slightly unnatural but do not impede understanding.
        \item \emph{2 points:} There are noticeable grammatical, spelling, or structural issues that affect fluency, but the overall message is understandable.
        \item \emph{1 point:} Many errors in grammar, spelling, punctuation, or word choice, affecting overall understanding and the fluency of the response.
    \end{itemize}

    \item \textbf{Coherence (1--4 points):} This dimension aligns with structure and coherence in the DUC quality issues, where the answer should be well-structured and organized. The answer should not merely be a compilation of related information but should develop step-by-step into a coherent body of information on a specific topic.
    \begin{itemize}
        \item \emph{4 points:} Clear structure and logical organization, with natural and smooth transitions between sentences. Not only is the information relevant, but it also demonstrates overall coherence of the response, avoiding harsh transitions or topic interruptions.
        \item \emph{3 points:} Generally clear structure and organization, though occasional logical jumps or transitions may not be natural. The thematic development is mostly complete, but the connection of some content is slightly mechanical or abrupt, slightly diminishing the overall coherence.
        \item \emph{2 points:} Clear breaks or jumps between sentences, with rather loose logical organization. The thematic development of the information is unclear, with some content appearing irrelevant or repetitive, making the overall presentation disjointed.
        \item \emph{1 point:} Sentences lack a clear logical relationship, with chaotic content organization, making it difficult to form a coherent theme development. There is a significant feeling of information piling up, lacking structural organization, making it difficult for the reader to understand the core idea.
    \end{itemize}
    
    \item \textbf{Naturalness (1--4 points):} Assesses whether the answer conforms to the habits of natural language expression, resembling the tone and logic of human everyday communication. Focus on word choice, tone, sentence structure, and the overall reading experience.
    \begin{itemize}
        \item \emph{4 points:} The expression is smooth, the tone is natural and close to everyday conversation, sentence structures are flexible and varied, with no stiffness. Word choice is precise and engaging, creating a sense of realism in the dialogue.
        \item \emph{3 points:} Overall, the expression is natural, but some parts may appear formulaic or mechanized. The tone is slightly flat but does not hinder understanding. Word choice is appropriate but may lack variety.
        \item \emph{2 points:} Language expression is partly stiff or awkward, the tone is unnatural, appearing mechanical or overly formal. There may be sentence structures or logical arrangements that do not conform to everyday language habits.
        \item \emph{1 point:} The mode of expression is clearly unnatural, the tone is stiff or stilted, completely detached from everyday language habits, difficult for people to understand or accept.
    \end{itemize}

    \item \textbf{Empathy (1--4 points):} Comprehensive assessment of an individual's ability to understand, empathize with, and respond to others' emotions.
    \begin{itemize}
        \item \emph{4 points:} Able to deeply understand and empathize with others' emotions, responding appropriately to others' feelings, demonstrating high emotional sensitivity and care.
        \item \emph{3 points:} Can understand others' emotions and appropriately respond to most situations, but may occasionally fall short in complex or subtle emotional expressions.
        \item \emph{2 points:} Sometimes understands others' emotions, but responses may lack depth or contain misunderstandings, showing some emotional detachment.
        \item \emph{1 point:} Struggles to understand others' emotions or emotional reactions, lacks empathy, responses often seem inappropriate or irrelevant.
    \end{itemize}

    \item \textbf{Patient-Centricity (1--4 points):} Comprehensive assessment of an individual's ability to center communication on the patient's needs, providing clear, practical, and approachable information in a conversational manner.
    \begin{itemize}
        \item \emph{4 points:} Communication is highly patient-centered, using concise and approachable language, delivering practical advice tailored to the patient's situation, and maintaining a logical and conversational tone that fosters trust and understanding. Without listing ideas.
        \item \emph{3 points:} Generally patient-centered, but may lack full clarity, practical detail, or conversational ease in some parts while still being largely useful and understandable.
        \item \emph{2 points:} Moderately patient-centered but may include excessive technical details, lack practical guidance, or have a less approachable tone, making it harder for patients to follow or apply.
        \item \emph{1 point:} Communication fails to center on the patient, being overly formal, academic, or irrelevant, and lacking clarity, practicality, or approachability.
    \end{itemize}
\end{itemize}

\section{Pragmatic Rules}
\label{sec:appendixrules}

Our pragmatic rule acquisition followed a rigorous and multi-phase process. First, we conducted a systematic review of medical literature. Second, we summarized the principles that doctors should follow in communication with patients into three aspects: what format to use for replies, what information to convey, and how the user will feel. Third, we distilled actionable rules from these principles. The pragmatic rules used in this paper are listed below.

\begin{itemize}
\item Without listing ideas, reply like a human being\cite{longwen:how}.
\item The reply is easy to understand, and the technical terms have relevant explanations\cite{Katherine:jargon}.
\item Showing empathy: a comprehensive assessment of an individual's ability to understand, empathize with, and respond to others' emotions\cite{xiaoyu:physicianempathy}.
\item The reply has authenticity and feasibility\cite{CHEN2020102253}.
\item Clear structure and logical organization, with natural and smooth transitions between sentences\cite{JIANG2024108161}.
\end{itemize}

\section{LAPI vs. CoT Variants Performance with GPT-4}
\label{sec:cotvar}
To demonstrate the distinct advantages of the LAPI framework over Chain-of-Thought (CoT) prompting, particularly in providing actionable advice, we conducted a comparative analysis between LAPI and several enhanced CoT variants. This evaluation aims to verify whether LAPI's superior performance persists even when CoT prompts are explicitly designed to align more closely with LAPI’s approach, thus clarifying LAPI’s technical contribution. Specifically, we compared LAPI against four CoT variants using GPT-4: (1) Baseline CoT with \textit{''Let’s think step by step''} prompting; (2) CoT-(P.R.), which incorporates pragmatic rules; (3) CoT-(HBM), which integrates HBM theory; and (4) CoT-(HBM+P.R.), which adds pragmatic rules to CoT-(HBM). As shown in Table 3, LAPI consistently outperforms all CoT variants across multiple metrics, reinforcing its distinct technical advantages.

\begin{table}[h]
\small
\centering
\label{tab:ablation2}
\resizebox{\columnwidth}{!}{
\renewcommand{\arraystretch}{1}
\begin{tabular}{l|ccccccc}
\toprule
\textbf{Approach} & \textbf{Flu.} & \textbf{Coh.} & \textbf{Nat.} & \textbf{Emp.} & \textbf{Pat-C.} & \textbf{Rec.} & \textbf{Pre.} \\ \midrule
CoT & 4.00 & 3.70 & 3.10 & 2.35 & 2.85 & 0.37 & 0.08 \\
CoT-(P.R.) & 4.00 & 3.90 & 3.75 & 3.65 & 3.55 & 0.36 & 0.07 \\
CoT-(HBM) & 4.00 & 3.85 & 3.15 & 2.80 & 3.05 & 0.36 & 0.08 \\
CoT-(HBM+P.R.) & 4.00 & 3.80 & 3.90 & 3.60 & 3.75 & 0.35 & 0.07 \\
LAPI & \textbf{4.00} & \textbf{3.90} & \textbf{3.95} & \textbf{3.85} & \textbf{3.80} & \textbf{0.38} & \textbf{0.07} \\
\bottomrule
\end{tabular}}
\caption{Performance comparison between LAPI and improved CoT prompts using GPT-4.}
\end{table}

\end{document}